\documentclass{iaus}
\usepackage{graphicx}

\title[Methanol masers and high-mass star formation] 
{Methanol masers as tools to study high-mass star formation}

\author[Pestalozzi, M.R.]   
{Michele R. Pestalozzi$^1$}

\affiliation{$^1$Centre for Astrophysics Research, University of
  Herftodshire, AL10 9AB Hatfield, UK \break email:
  michele.pestalozzi@gmail.com
}

\pubyear{2007}
\volume{242}  
\pagerange{1-8}
\date{March 2007}
\jname{Astrophysical masers and their environment}
\editors{J. Chapman \& W. Baan, eds.}

\begin{document}

\maketitle

\begin{abstract}
In this contribution I will attempt to show that the study of galactic
6.7 and 12.2\,GHz methanol
masers themselves, as opposed to the use of methanol masers as
signposts, can yield important conclusions contributing to 
the understanding of high-mass star formation. Due to their exclusive
association with star formation, methanol masers are the best tools to
do this, and their large number allows to probe the entire Galaxy. In
particular I will focus on the determination of the luminosity
function of methanol masers and on the determination of an unambiguous
signature for a circumstellar masing disc seen edge-on. Finally I will
try to point out some future fields of research in the study of
methanol masers. 

\end{abstract}

\section{Introduction}

The galactic maser emission of methanol appears to be exclusively
associated with sites of high-mass star formation (\cite{min03}). In
particular, the 
{\it class~II} masers at 6.7 and 12.2\,GHz are among the brightest
masers in the Milky Way. Since their discovery, a number of
extended searches have been put into action with the aim of gathering
a potentially large number of objects to be studied in depth. A good
list of these searches is presented in \cite{pes05}. Because
of their high brightness, methanol masers are obvious targets for
interferometric observations 
that reveal the finest dynamical and morphological details at large
distances (1\,AU at 1\,kpc).  

\begin{figure}
\begin{center}
\includegraphics[width=5cm, angle=90]{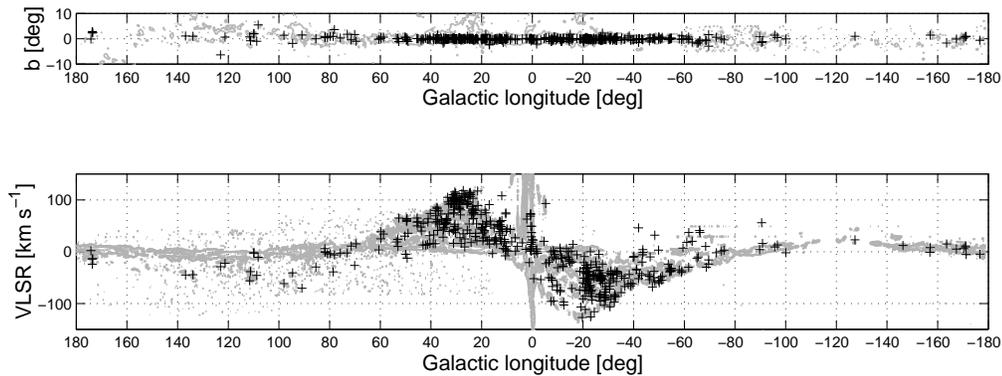}
\caption{Distribution of methanol masers in Galaxy, superposed on the CO
  contours from \cite{dam87}, in space (top) and
  LOS velocity 
  (bottom). The methanol masers seem to accurately follow the overall
  structure of the Galaxy, both in space and LOS
  velocity. Particularly visible in the bottom panel is the fact that
  methanol masers are tracing the spiral arms ($150^{\circ} > l >
  80^{\circ}$ and $-40^{\circ} > l > -90^{\circ}$) and the high
  rotational velocity of the nuclear ring ($l \approx 0^{\circ}$).}
\label{fig:methall}
\end{center}
\end{figure}

The census of 6.7\,GHz masers in the Milky Way counts to date some 520
sources, all potentially locating a high-mass protostar
(\cite{pes05}). This number is subject to increase as the Methanol
MultiBeam (MMB) Survey (see contribution by J. Green) covers more
regions of the galactic plane. The distribution of the known methanol
masers in the Galaxy is shown in Fig.~\ref{fig:methall}.

Research connected with methanol masers can be summarised and organised
as shown in Table~\ref{tab:maser_stud}. In the squares, I put some examples
of the type of investigation that were undertaken to study methanol
masers. Methanol maser 
studies can be divided into two main categories: studies having the
maser emission itself as target of their investigation on the one hand, 
and studies having the source hosting the maser emission as their target on
the other. Both these paths give rise to two subclasses: studies
of the maser sources as a group (statistical or global analysis) on
the one hand, and deep studies on specific sources on the other. The
outcomes from these studies can also be distributed into a similar
diagram. The goal is to make the outcomes fall in the most general
possible category (bottom left square in Table \ref{tab:maser_stud}),
i.e. understand high-mass star formation.

The common property of the early searches for 6.7\,GHz (and
12.2\,GHz) methanol masers was that they concentrated their efforts in the
characterisation of the host of the masers (bottom left square of
Table \ref{tab:maser_stud}). This was in fact a natural consequence of
the strategy adopted to find new sources, i.e. targetting 
IRAS sources according to selection criteria corresponding to
particular objects, as Ultra Compact (UC) HII regions with more or
less red colours (e.g. \cite{schu93},\cite{szy00}), or known star
formation regions hosting OH masers (as e.g. \cite{cas95}). The
relatively low detection rate of these surveys revealed the inaccuracy
of the definition of methanol maser hosts and called for untargetted
searches, where large areas of sky were covered  
 with observations (e.g. \cite{elli96}, \cite{szy02},
\cite{pes02b}, \cite{pan07}). The main outcome of these studies was
that a large 
fraction of methanol masers at 6.7\,GHz were found not to be associated with
known bright IR emitters but with very deeply embedded and therefore
probably young objects (e.g. \cite{pes02a}). This conclusion relies on accurate
astrometric information both for the masers and for the IR objects,
which is only recently starting to become available at a large number
of wavelengths and for a large number of targets. More recently, large
follow-up campaigns in the millimetre and 
sub-millimetre bands in continuum and spectral line emission have
given further insight into the nature of the 
hosts of 6.7\,GHz methanol masers (\cite{pur06},
\cite{hil05}). The main outcome from these studies
is that methanol masers are probably the first signal for the
existence of a high-mass protostar when it is still embedded in its
thick primordial dust cocoon. 
 
Studies having the masers themselves as object are more challenging. This
is mainly due to the fact that the maser emission is non-thermal in
nature ad therefore the extraction of information on the physical conditions
of the emitting medium is not trivial. Theoretical modelling of the
maser emission of methanol faces the challenge of including a larger
number of molecular transitions in the calculations, potentially all
interacting and producing a large number of masing transitions
(e.g. \cite{sob97}, \cite{cra05}). These studies are tightly linked to
multi transitional observations of maser-bearing objects, that are
starting to include larger samples of targets (see \cite{pes05} for a
list). In the near future these studies will greatly benefit of the
clear improvement of backends that will allow simultaneous observation
of different frequency bands covering large regions of the spectrum at
once. Polarisation measurements appear to be technically challenging,
and from the levels of polarisation obtained, it is not always easy to
discriminate between pumping effects and scattering effects
(\cite{wie04}). Finally, the quality of the conclusions 
from variability studies is critically dependent on long time
monitoring, and has only recently started to produce interesting
results. Some sources show a remarkable regularity, and represent a
challenge for modelling (see e.g. \cite{goe04}).  

\begin{table}
\begin{center}
\caption{Conceptual division of the study of methanol masers.}
\label{tab:maser_stud}
\begin{tabular}{l|l|l}\hline
 & Global studies & Particular studies \\
\hline
Masers & maser luminosity & NGC\,7538: disc signature \\
 & structure of the Galaxy & Location of new sources \\
 & maser theory & proper motions \\
 & polarisation & (variability) \\
\hline
Hosts of masers & Associations & IRAS 20126, S255 \\
 & SEDs & Protoclusters \\
 & follow-up observations & \\
\hline
\end{tabular}
\end{center}
\end{table}

In this contribution I will concentrate on the upper row of Table
\ref{tab:maser_stud} and present one example for each square. In
particular I will concentrate in the extraction of the luminosity
function of the masers and on the recognition of a unique signature
for an edge-on rotating disc marked by maser emission.

\section{Global study: the luminosity function of
  the masers} 

The distribution of the known methanol masers in the Milky Way is
shown in Fig.~\ref{fig:methall}. The majority of the masers are located
in a ring, covering galactic longitudes of $\pm$50$^{\circ}$. What
is interesting is that a comparison of the galactic distribution of
methanol masers with OB associations (\cite{bro00}) shows excellent
agreement. This fact reinforces the 
association of methanol masers with high-mass star formation. 

The study of the luminosity function of astronomical objects has to
first solve the problem of the distance determination. In the Milky
Way, kinematic distances can be attributed to sources observed in
spectral line emission (as e.g. masers). Kinematic distances for
sources orbiting the Galactic centre on orbits internal to the Sun's
suffer from an ambiguity which can be resolved only on a case to case
basis. Several methods have been used to solve this ambiguity with
varying success (see e.g. \cite{bus06} for the HI self absorption
method, and \cite{vdw05} for a probabilistic approach). Here I present
an approach to the study of the luminosity function of methanol masers
that removes the problem of the heliocentric distance determination. 

\begin{figure*}
\includegraphics[width=6.7cm]{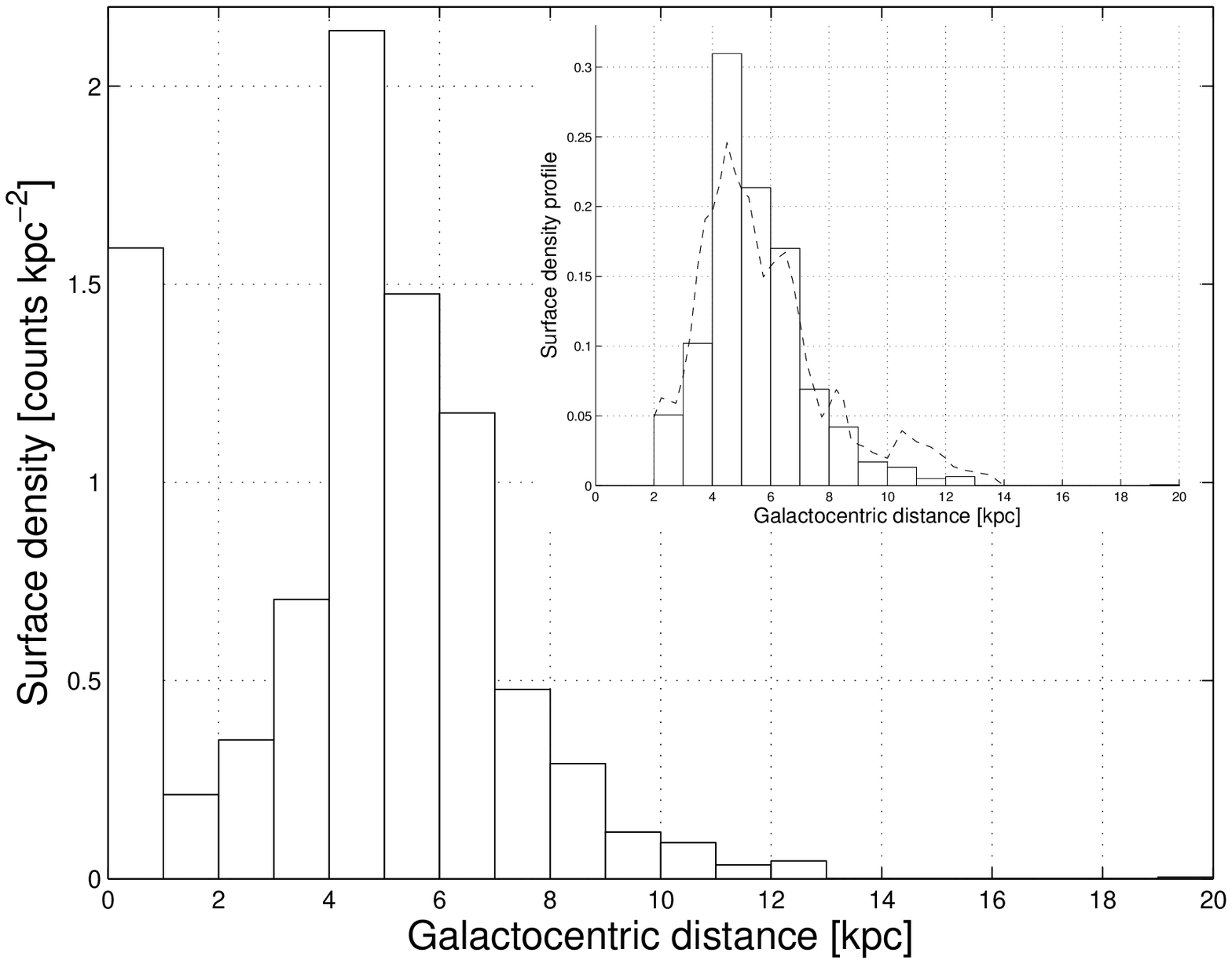}\includegraphics[width=6.7cm]{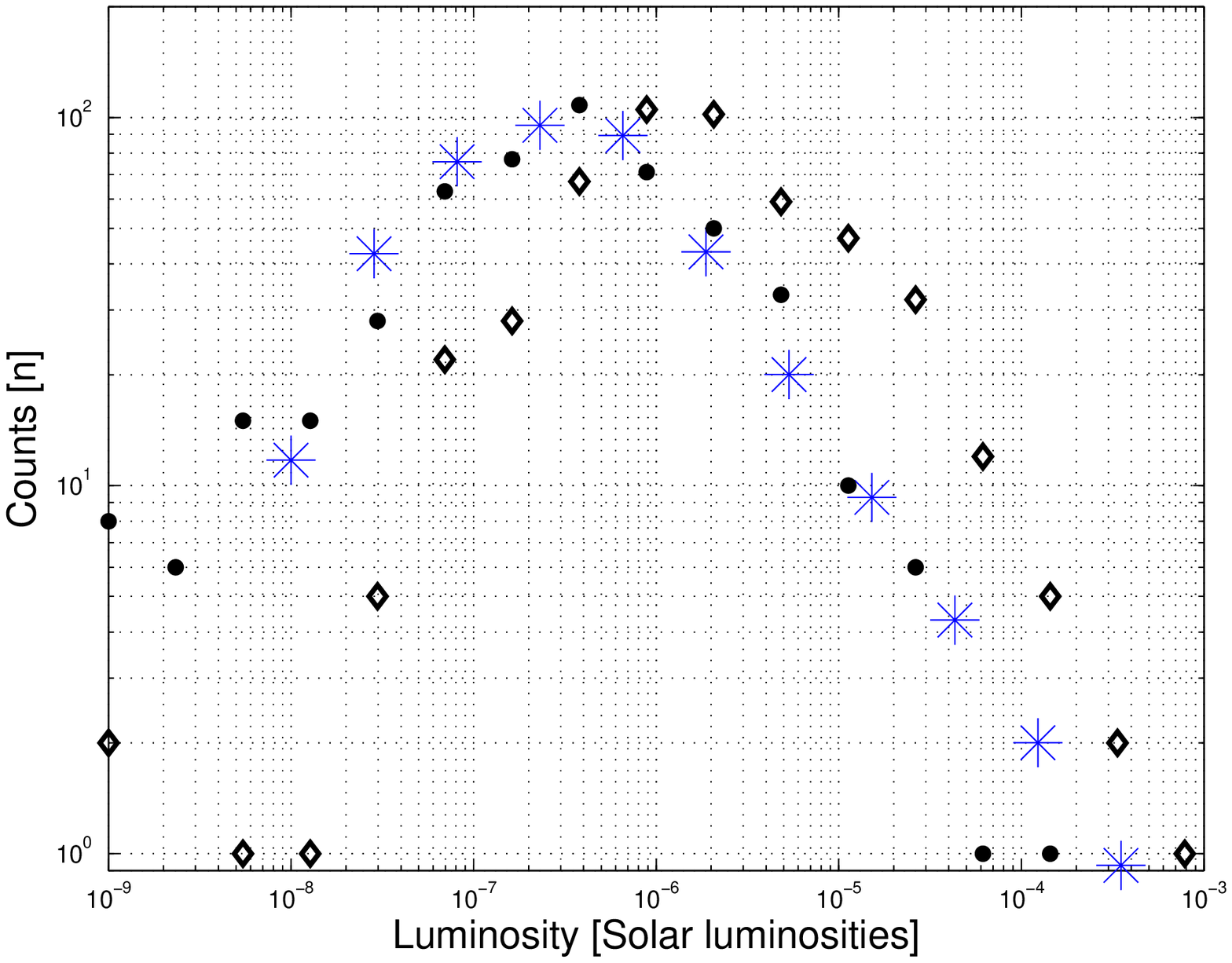}
\caption{Left: Surface density of the known methanol masers in galactocentric
  distance. The distribution shows a peak at the same radius than OB
  associations. The inset shows a comparison of the surface density of
  the masers with the molecular gas in the Milky Way. Right:
  Luminosity distribution of methanol masers in the Galaxy. Diamonds
  and circles plot all the (ambiguous) sources at the near- and far 
  heliocentric distance, respectively. Stars show a synthetic
  luminosity distribution obtained by observing with a 2\,Jy
  sensitivity a population of 5000 masers spatially distributed
  according to a Gaussian and having an intrinsic luminosity
  distribution as a power-law between cutoffs of index -1.7.}
\label{fig:gc_masers}
\end{figure*}

What is needed is to disentangle the spatial distribution of the
sources under investigation from their intrinsic luminosity
distribution. In fact, the {\it observed} luminosity distribution of
masers (diamonds and circles in Fig. \ref{fig:gc_masers}) is the
convolution of a spatial distribution with an {\it intrinsic}
luminosity distribution after applying an observational sensitivity
cutoff. Knowing the spatial distribution of the masers and the
sensitivity cutoff leaves the intrinsic luminosity function to be the
only unknown in the problem.  

The starting point is the distribution of masers in galactocentric
distance. The latter is not affected by any ambiguity, and only assumes a
rotation curve of the Galaxy. The inaccuracies of the rotation curve
of the Galaxy do not significantly affect the final result,
because, as it is seen in Fig. \ref{fig:gc_masers}, most of the
methanol masers are distributed in a ring with galactocentric radius
$\sim$5\,kpc: at that radius, even the influence of the central bar is
not strong enough to produce significant deviations from a circular
motion (see e.g. \cite{fux99}). The distribution of masers in
Fig.\ref{fig:gc_masers} is assumed to represent with a high level of
confidence {\it the shape} of the true, azimuthally averaged, spatial
distribution of methanol masers in the Galaxy, $F(R)$. This will be
convolved with the intrinsic luminosity distribution $G(L)$ and
``observed'' with a certain sensitivity to fit the galactocentric
distribution of observed masers shown in Fig.\ref{fig:gc_masers}. 

The free parameters in the fit are the real total number of masers in the
Galaxy (the integral of $F(R)$) and the function $G(L)$. To keep the
number of unknowns to a minimum $G(L)$ was expressed as a single
power-law between two cutoffs. These 
cutoffs were fixed by taking a range of sensible luminosities from the
literature, so that the only two free parameters in the fit were the
index of the luminosity function and the total number of methanol masers in the
Galaxy. The two parameters cannot be found simultaneously but a
plausible range can be estimated. On the basis of previous
estimates of the total number of sources in the Galaxy (\cite[van der
  Walt 2005]{vdw05}), the index for the luminosity function was found
to lie between -1.5 and -2.0. In Fig.\ref{fig:gc_masers} an example of
observed synthetic luminosity distribution is shown (stars). The model
consists of a total number of 5000 sources 
distributed in luminosity according to a power-law with index -1.7 and 
observed with a detection threshold of 2\,Jy. The details of this work
are presented in \cite{pes07a}.

The significance of the intrinsic luminosity distribution of the
 masers is still not clear, both in terms of high-mass star formation
 and in terms of maser modelling. A major contribution
will be given by the completion of the MMB Survey, that is expected to
determine with a very high level of confidence how many methanol
masers there are in the Milky Way and thus limiting one more parameter
in the model presented here.

\section{Particular study: the disc signature}

\begin{figure}
\includegraphics[width=7cm]{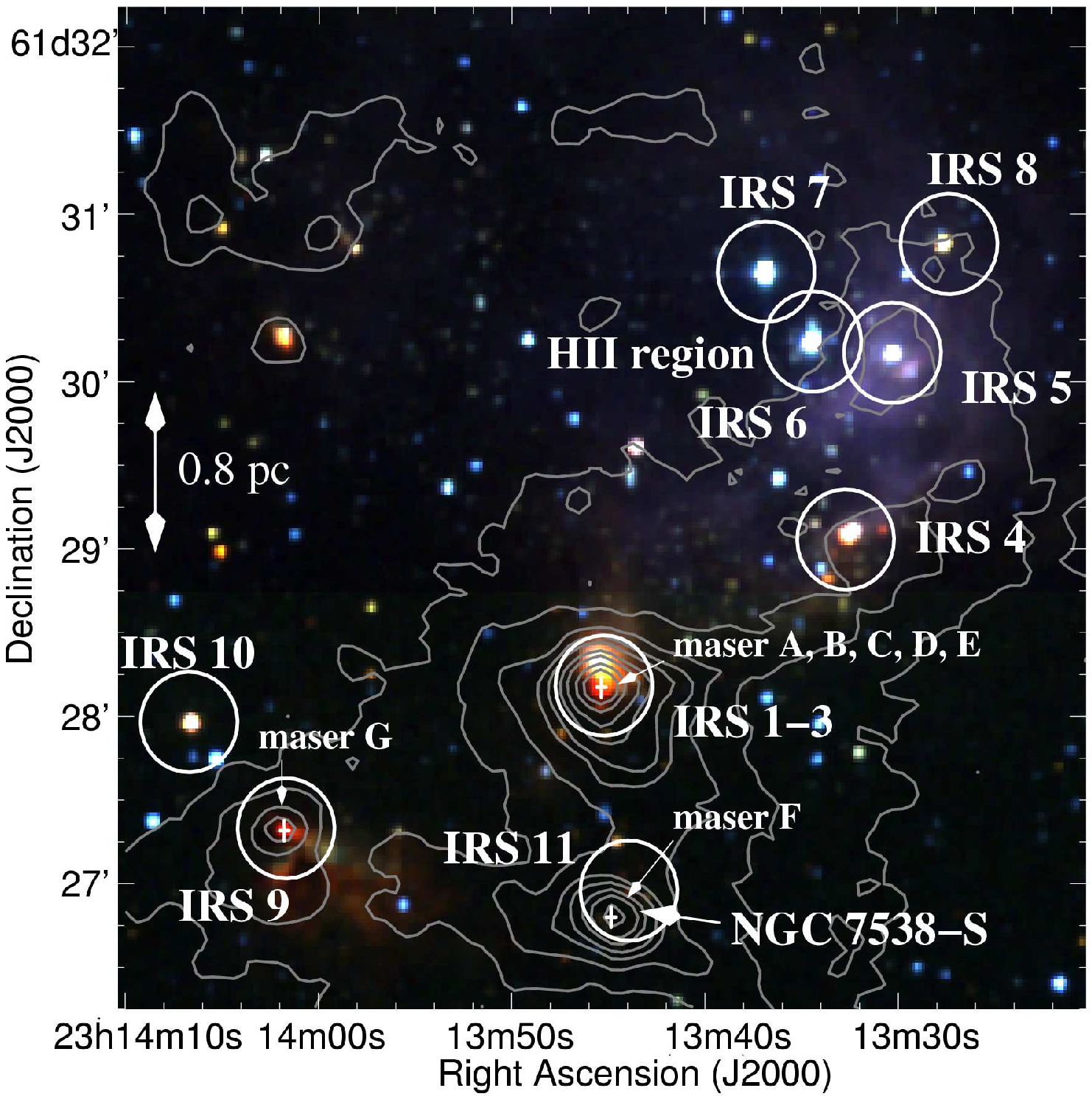}\includegraphics[width=6cm]{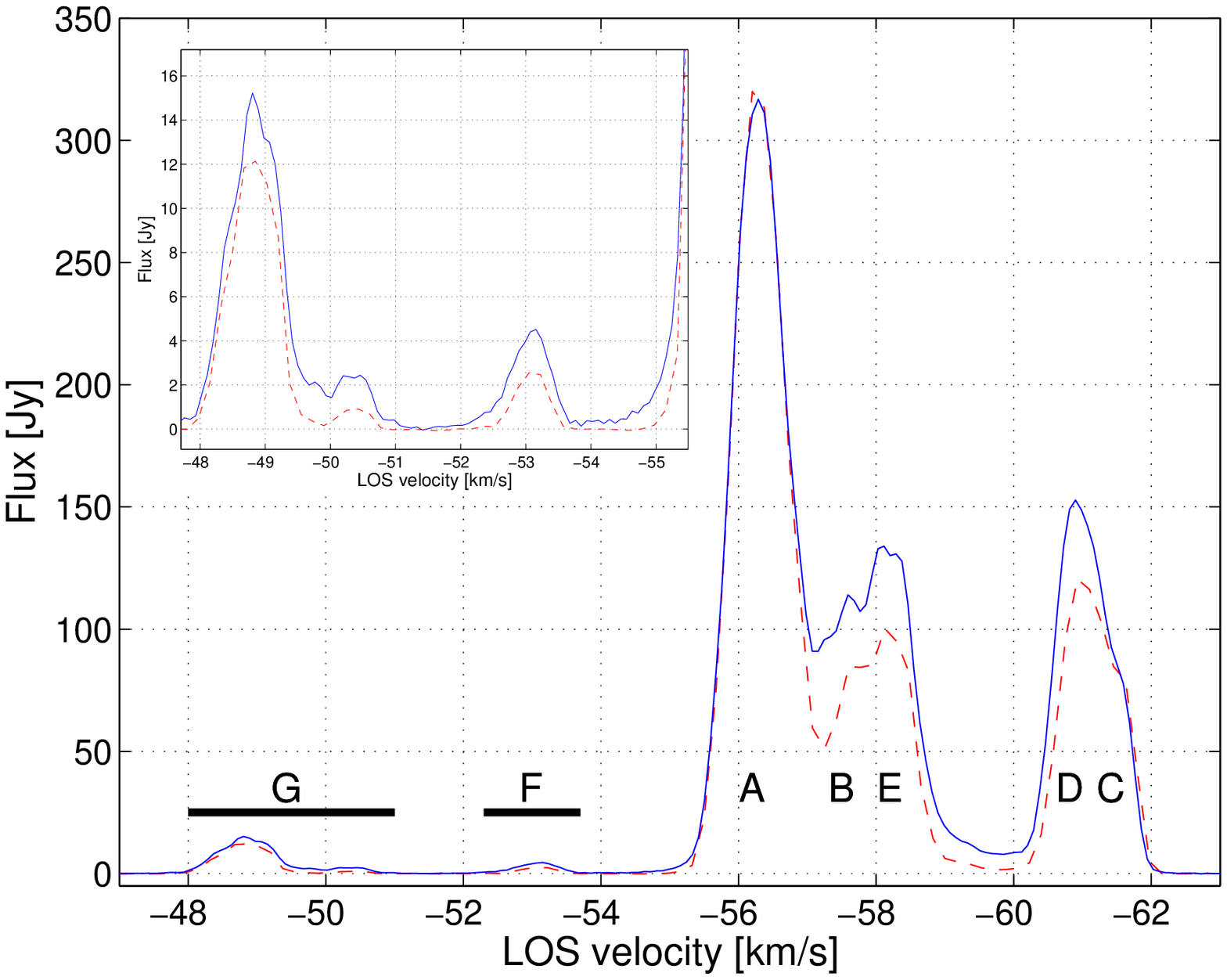}
\caption{Left: General view of the high-mass star forming region
  NGC7538. Grey contours of the 1.2\,mm dust continuum are overlaid
  on a JHK 2-mass image. Circles indicate the catalogued IR sources,
  crosses the methanol masers. The methanol masers
  appear in association with the youngest and most massive
  objects. Right: Spectrum of the 6.7\,GHz maser emission toward
  NGC7538. Features {\sf A, B, C, D, E} are cospatial with IRS~1. The
  maser disc is feature {\sf A}. The inset is a zoom of the features
  {\sf F} and {\sf G}. The 12.2\,GHz spectrum is similar to
  the 6.7\,GHz one.}
\label{fig:ngc_all}
\end{figure}

NGC7538 is a very well known and deeply studied high-mass star
forming region in in the Perseus arm. It contains some 11
catalogued IR sources spanning over a wide range of ages. The centre
of the figure is occupied by an UC HII region (IRS\,1) that hosts
bright 6.7 and 12.2\,GHz methanol masers (see 6.7\,GHz spectrum in
Fig. \ref{fig:ngc_all}).  

All spectral features have been observed at high angular resolution
using MERLIN and the EVN. In particular, features {\sf F} and {\sf G} were
recently positioned and appear to be cospatial with the youngest and
most massive sources in the region, namely NGC7538~S and IRS~9,
respectively (\cite{pes06}). By determining the position of the masers
with an accuracy of better than 30\,mas, the authors claim 
to have accurately determined the position of the protostars. 

\begin{figure*}
\includegraphics[width=6.5cm]{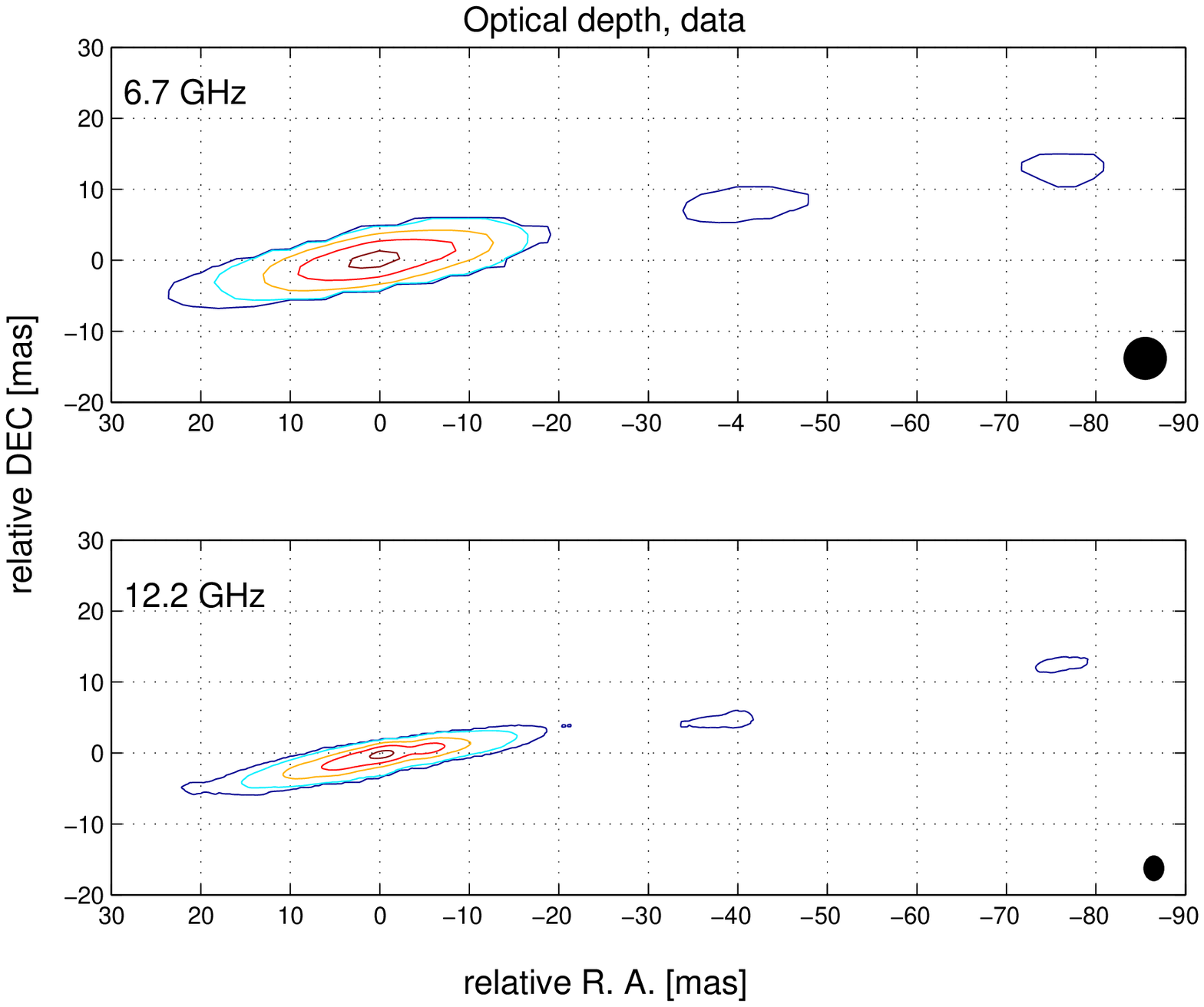}\includegraphics[width=6.5cm]{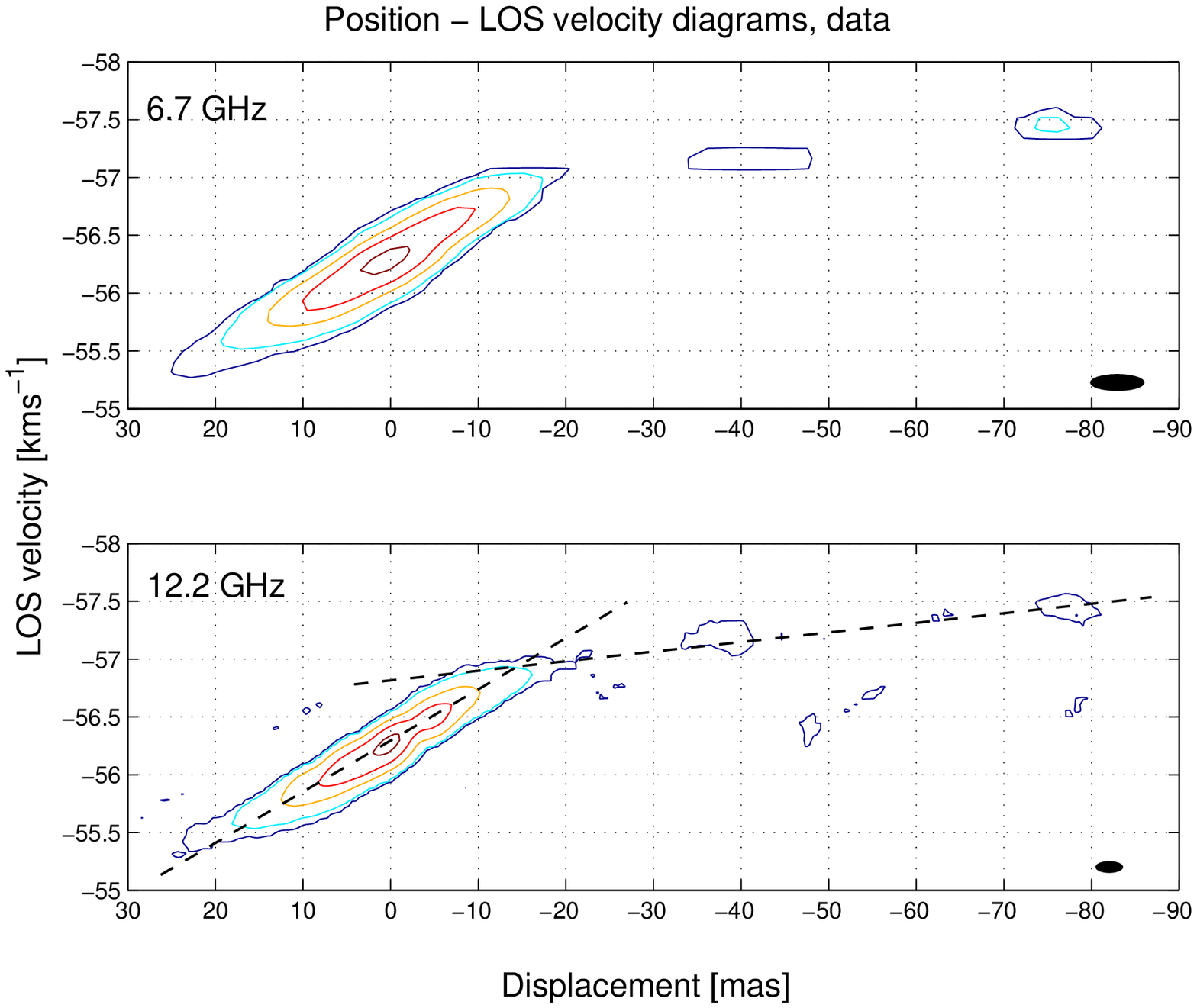} 
\caption{Maps (left) and velocity-displacement diagrams (right) of the
  main spectral 
  feature of the 6.7 and 12.2\,GHz methanol masers toward NGC7538
  IRS~1. The beam size is shown in the bottom right corner in all
  panels. The contours are 1, 5, 10, 30, 50, 70, 90\% of the peak in
  all panels. Dashed lines indicate the change in gradient between the
  main feature and the outliers.}
\label{fig:main_alldata}
\end{figure*}

The main spectral feature ({\sf A} in Fig. \ref{fig:ngc_all}) reveals
a single, smooth spatial feature 6-20 beams across and shows a mainly
linear LOS velocity gradient (Fig. \ref{fig:main_alldata}).

The linear structure in both space and LOS velocity was naturally
interpreted as arising from a (Keplerian) rotating disc seen edge-on
amplifying a background continuum, and where the {\it maser spots} are
orbiting a central object at a fixed radius (\cite{min98}). The 
maser emission was subsequently modelled as coming from a uniformly
distributed material in a rotating disc seen edge-on and confined
between an inner and outer radii. This avoided the assumption of a
single radius and produced astonishingly accurate fits to the data
(\cite{pes04a}).  

Assuming amplification of a background continuum $I_B$, the optical depth of
the maser emission at displacement $\theta_x$ and $\theta_y$ from the
centre of each frequency plane with LOS velocity $v$ is expressed by
$\tau(\theta_x,\theta_y,v) = \ln(I(\theta_x,\theta_y,v)/I_B)$. 
In the case of a thin disc, one spatial dimension can be dropped and
the modelling reduces to the study of the position-LOS velocity
diagram of the normalised optical depth $F(\theta_x,v) =
\tau(\theta_x,v)/\tau_0$, where $\tau_0$ is the optical depth at
displacement 0. The function $F$ is then expressed as:
\begin{equation}
F(\theta_x,v) = \int \eta(\rho) \exp\left[-\frac{1}{2}
  \left(\frac{v-\Omega(\rho)\,\theta_x}{\Delta
  v_D}\right)^2\right]\frac{{\rm d}\rho}{\sqrt{1-(\theta_x/\rho)^2}} 
\label{eq:fund_rot}
\end{equation}
where $\Omega(\rho)$ is the angular velocity at every radius and the
product $\Omega(\rho)\,\theta_x$ is the LOS component of the rotation
velocity. $\Delta v_D$ is the linewidth at $\theta_x=0$ and $\eta$ is
the gain of the maser that also defines the limits of
integration. \cite{pes04a} modelled
$\eta$ as a power-law with index 
0.5 and with sharp cutoffs at inner and outer radii $\rho_{in}$ and
$\rho_{out}$.  

The aim is to reproduce the right panels in
Fig.~\ref{fig:main_alldata}. Two features are of extreme importance in
this matter: the fact that most of the emission is symmetric around a
peak (that we will consider to be the central LOS across the disc) and
that there is a significant drop of the overall gradient at
displacements larger than 10-15\,mas. These two facts greatly
constrain the dynamics in the modelling. 

The study of $F$ is best summarised by the
study of the {\it spine}, defined by the maximum of $F$ at every LOS,
$\partial F / \partial v =0$. 
In the case $\Omega(\rho)=constant$, i.e. solid body rotation, the
study of the spine becomes very much simplified, as the exponential in 
eq.~\ref{eq:fund_rot} comes out of the integral. The location of the
spine is defined by $v-\Omega\,\theta_x =0$ and the optical depth
on the spine by the integral. The spine is then a straight
line with constant slope ($\Omega$). $F$ on the spine will have maxima
at displacements tangent to the inner edge of the disc and a local
minimum at $\theta_x=0$. The spectrum will have two features at $v=\pm
\Omega \theta_{in}$, where $\theta_{in}$ is the LOS tangent to the
inner edge. In the special case of $\Omega=0$ (quiescent
disc) the spine will be parallel to the displacement axis, $F$ on the
spine will be the same as the previous case and the spectrum will have
only one feature, centered at $v=0$ and of width $\Delta v_D$. The
above considerations unambiguously show that solid body rotation (and
a quiescent disc) cannot reproduce the data in
Fig.~\ref{fig:main_alldata}, differential rotation is required. 

The study of the spine for $\Omega \ne const$ gives a constraint on the 
rotation velocity: in order for $F$ to have a maximum at $\theta_x=0$,
the rotation velocity at the inner edge has to be at least some 3-4
times higher than $\Delta v_D$ (\cite{pes07b}). Also, in order to
account for the drop in gradient (dashed lines in the lower right
panel of Fig.~\ref{fig:main_alldata}), the spine analysis indicates
that the dynamics has to be Keplerian, i.e. $\Omega \sim \rho^{-3/2}$.
Finally, it is important to notice that if the 
rotation velocity does not meet the above constraint, both spectrum
and $F$-profiles ($F$ on the spine) will have three peaks of emission,
in the same way that was predicted by e.g. \cite{pon94}. 

The general conclusion at this point is that a  maximum of emission in
the centre of 
both map and velocity-displacement diagrams is a strong hint for
(fast) differential rotation seen edge-on. This is also supported by
the existence of a clear bend of the gradient in the velocity-displacement
diagram. These two facts unambiguously exclude solid body rotation
(and a quiescent disc) as these would produce linear features both in
space and velocity but no maximum in the centre, and no bend in the
velocity diagram.

\section{Summary and future views}

In this paper I tried to show how the study of methanol masers
themselves (as opposed to the study of the environment of methanol masers)
can lead to important conclusions about the maser emission 
(luminosity function) and the hosts of the masers (disc signature and
accurate localisation of high-mass protostars). 

The luminosity function of the maser will be refined in the near
future thanks to the outcomes of the MMB Survey, which will also yield
the ultimate census of the  
population of methanol masers in the Galaxy. The study of the accurate
luminosity function of the masers will give insights into some
characteristics of the maser mechanism, instrumental for 
extracting the physical conditions of the maser environment from the
maser emission itself.

By accurately modelling the maser emission in one source, we were able
to unambiguously determine what the signature for a circumstellar
(differentially) rotating disc is as opposed to e.g. a bipolar
outflow. Such model will be applied to other sources in order to test
the existence of thin discs around high-mass protostars. The finding
(or not!) of further discs around high-mass protostars will contribute
in refining the high-mass star formation scenario.

Furthermore, there is a strong call for studies of the Galactic structure of
the Milky Way, as methanol masers seem to represent fairly equally all
galactic longitudes (and latitudes) and carry an important dynamical
information. This is unprecedented, as previous studies of the
structure of our Galaxy were relying on the observation and
positioning of HII regions. 

One aspect that will be crucial for the future studies of methanol
masers will be the gathering of data on the extended maser
emission. As it has been observed by \cite{min02} and
\cite{harv06}, it appears that maser emission
can probe large surroundings of a high-mass star forming region. The
mechanisms of this low brightness maser emission will have
to be understood in order to be able to correctly interpret the data
and increase our understanding of the physical conditions around
high-mass protostars. 

Finally it is exciting to keep in mind that the next few years will be
characterised by large scale galactic legacy surveys at a number of
wavelengths. In particular, the SCUBA-2 surveys (SASSy and JPS) will
yield the most complete catalogue of star forming clouds in the Milky
Way detected at sub millimetre wavelengths. The cross correlation of
these catalogues with the final MMB 
catalogue will enable the definition of an evolutionary sequence for the
initial stages of the formation of high-mass stars. More in the future,
the Herschel survey of the Galactic Plane will give the long waited
view of the Galaxy at previously unavailable 
frequencies. Again, this will allow an accurate characterisation of methanol
maser hosts and ultimately the definition of an accurate evolutionary
sequence for high-mass star formation.

\begin{acknowledgments}
The present paper is a summary of the work done in collaboration
with an extended group of people to whom I am deeply thankful:
A. Chrysostomou and J. Collett (Univ. of 
Hertfordshire, UK), M. Elitzur (Univ. of Kentucky, USA), V. Minier (Saclay,
Paris, France), J. Conway (Onsala Space Observatory, Sweden), R. Booth
(HartRAO, South Africa), and the NGC7538 collaboration \\ 
(see the website {\sf
  http://star-www.herts.ac.uk/$\sim$michele/website\_ngc/index.htm}). 
I also thank D. Nutter for the careful reading of the final
manuscript. 

\end{acknowledgments}

\begin{discussion}

\discuss{Y. Rodriguez}{What is the nature of the outliers in your maps
  of the NGC7538 maser emission at displacements $>$30\,mas?}

\discuss{Pestalozzi}{These are considered to be local enhancements of
  the optical depth that do not fall on the spine. They helped in
  constraining the drop of gradient that finally indicated Keplerian
  rotation.}

\end{discussion}

\end{document}